\documentclass[a4paper,12pt]{article}

\usepackage[utf8]{inputenc}
\usepackage{epsfig}
\usepackage{graphicx}
\usepackage{amsmath}
\usepackage{amssymb}
\usepackage{cmap}
\usepackage{bm}
\usepackage{hyperref}
\usepackage{array}
\usepackage{epsfig}
\usepackage{array}
\usepackage{color}
\usepackage{latexsym}
\usepackage{xypic}
\usepackage{amscd}

\usepackage{cite} 

\usepackage[left=1.0in, right=1.0in, top=1.0in, bottom=1.0in]{geometry}

\newcommand{\version}{\sf arXiv:~v2,~~28-06-17.}
\newcommand{\thedate}{June 28, 2017}

\title{Color screening in flux tubes and in the color Coulomb potential from the QCD Field Correlators \thanks{\version}}

\author{
M.S.~Lukashov$^{a,b,}$\thanks{lukashov@phystech.edu}\,\, and Yu.A.~Simonov$^{a,}$\thanks{simonov@itep.ru}\, \\
\\
$^a$ \small{\em Alikhanov Institute for Theoretical and Experimental Physics,}\\
\small{\em B. Cheremushkinskya 25, 117218 Moscow, Russia}\vspace{0.25cm}\\ 
$^b$ \small{\em Moscow Institute of Physics and Technology,}\\
\small{\em Institutskiy per. 9, 141700 Dolgoprudny, Moscow Region, Russia}
}\bigskip
\date{\thedate}

\newcommand{\be}{\begin{equation}}
\newcommand{\ee}{\end{equation}}

\def\fun#1#2{\lower3.6pt\vbox{\baselineskip0pt\lineskip.9pt
\ialign{$\mathsurround=0pt#1\hfil ##\hfil$\crcr#2\crcr\sim\crcr}}}

\newcommand{{\SD}}{\rm SD}

\newcommand{\ver}{\mbox{\boldmath${\rm r}$}}

\newcommand{\veR}{\mbox{\boldmath${\rm R}$}}

\newcommand{\vek}{\mbox{\boldmath${\rm k}$}}

\newcommand{\ven}{\mbox{\boldmath${\rm n}$}}

\newcommand{\veE}{\mbox{\boldmath${\rm E}$}}

\newcommand{{\Mc}}{\mathcal{M}}

\newcommand{\lan}{\langle}
\newcommand{\ran}{\rangle}

\begin{document}
\maketitle

 
\begin{abstract}
{
\noindent{Colorelectric and Colormagnetic structure of the flux  tubes, connecting heavy
quark and antiquark, is investigated in the framework of the Field Correlator
method which  describes all  resulting fields in terms of correlators $D^E$ and
$ D^E_1$. The latter have  been  computed via gluelumps, which allows to
predict the resulting  distribution of color fields $\veE (\ver), $ and
colormagnetic  currents $\vek(\ver)$ in the flux tubes. It is shown, that at
large distances $r\gg \lambda \approx 0.2 $ fm the whole structure of fields
and relations between them  is similar to that of   the dual superconductor
theory, but  the basic dynamics, including small distances, is given by field
correlators of the real stochastic vacuum. The important contradiction between
the strong screening of color fields in the  width of flux tubes and
almost no screening in the perturbative $Q\bar Q$  potential is resolved.}\medskip

\rule{0mm}{5mm}

 \medskip

}
\end{abstract}

\newpage
\section{Introduction}

The flux tubes between heavy quark and antiquark are considered as a necessary
consequence of the color confinement mechanism, and were investigated
  numerically on the lattice during the last three decades,  see
e.g. \cite{1,2,3,4,5,6,7,8,9,10,11,12,13,14,15,16,17,18,19,20,21,22,23,24}.

It was understood that this physical phenomenon should exist, whatever is the
mechanism of color confinement, and hence only the detailed structure of flux tube
fields can distinguish between different models of confinement.

One of the most popular, however not derived from QCD,   is the model of dual
superconductor (MDS)  \cite{25,26}, where the QCD vacuum can be represented as
a coherent state of colomagnetic monopoles. The numerous studies, both in
theory and in numerical lattice works, have been  done for the last two
decades, trying to find the reasonable arguments and explicit formalism for
MDS, see the review papers
 \cite{27,28}
 and the references therein.

The difficulty of this approach is the lack of colormagnetic monopoles as real
physical objects or Euclidean solutions of QCD, so that one should consider
those as effective degrees of freedom in the real physical vacuum of QCD.

Meanwhile the  quantitative theory of color  confinement was suggested in
 \cite{29a,29b,29c}
 based on vacuum field correlators, developed in detail  for
application in hadron properties
 \cite{30a,30b}, including theory of Regge trajectories etc., the theory  of  chiral
symmetry breaking
 \cite{31},
 perturbation theory
  \cite{32}
 and the QCD thermodynamics
 \cite{33}.

Field correlators can be found from the solution for the gluelump Green's
functions  of   \cite{34,35a,35b}, which are connected back to field correlators in
a selfconsistent way \cite{36a,36b,36c}, which allows to define their properties for
distances $r>\lambda$, where $\lambda \approx 0.2$ fm is the vacuum correlation
length, $\lambda\sim 1/M_{\rm glp}, M_{\rm glp}$ is the lowest gluelump mass.

The problem of flux tubes in the framework of field correlator formalism was
discussed in the review papers \cite{37,38}, where it was shown, that the basic
properties of flux tubes are easily obtained from this formalism.

 Recently a
new formulation of the QCD equation of state and temperature transition was
accomplished \cite{39}, which required a very detailed structure of the
confinement dynamics, i.e. of the properties of field correlators, and those
can be tested in the flux structure.

The latter, as shown    below,  define all the fields in the flux tube, and
inversely, the study of flux tube fields gives information about details of
field correlators, i.e. the details of the confinement mechanism.

From this point of view the flux tubes are an important source of information
about the structure of confinement, including the temperature dependence of its
constructing details.

Recently new lattice measurements of flux  tube structure have been done in
\cite{40,41a,41b} and specifically in \cite{40} the first  accurate results have
been obtained both for $SU(3)$ and $2+1$ QCD. It is the purpose of the present
paper to compare these results with our analytic approach and to draw some
conclusions on the mechanism of flux tubes and confinement.

The extended study of  flux tubes in the framework of FCM was done in
\cite{38}, where the CE field distribution $\veE (\ver)$ was found in terms of
the confining correlator $D^E(x)$ and perturbative correlator $D_1^E(x)$,
yielding the standard picture of the QCD string between two sources $Q$ and
$\bar Q$. In addition an important step was done, defining the colormagnetic
current $\vek (\ver)$, which has the form of rings around the string, and it
was  also shown, that the ``dual'' London equation $ rot~ \vek = \lambda^{-2}
\veE (r)$ -- is satisfied at large  distances from the string axis, $r\gg
\lambda$.

This fact actually supports the idea, that the Field Correlator theory of
confinement at large distances to some  extent is equivalent  to the dual
superconductor picture, however  the former  allows to describe the flux tube
fields at all distances.

It is one of the aims of our paper, to go further in this direction and in
particular to present the distribution $|\vek (\ver)|$ as a function of $r$,
which can be further computed numerically on the lattice.

In addition, an interesting consequence of our theory is the distribution of
the CE field $\veE $, which is produced by the  color charges and screened in
the transverse, but not longitudinal  direction   as described by the
correlator $D^E_1$ and has no equivalent in MDS.

Indeed, the FC describe two kinds of CE fields, $E^{(1)}$ and $E^{(D)}$ due to correlators $D^E$ and $D^{E}_{1}$ respectively, and they have completely different distributions, in particular, $D^E$ gives the main body of the flux tube,while $D_1$ gives the screening of the color Coulomb interaction.

Actually in \cite{38}, the important problem of the screening of perturbative
fields due to  confinement was not fully investigated, and instead there was a
requirement of this screening at large distances, $r\gg \lambda$, where
$\lambda \approx 0.2$ fm is the vacuum correlation length. Below we give the
full answer to this problem of screening, based on the theory of confinement.

Another important development of the analysis of flux tube is its temperature
variation, which was done on the lattice in \cite{42}. This allows to measure
the $T$ dependence of correlators $D^E,D^E_1$, which plays the crucial role in
the temperature transition region, as shown in \cite{39}.

 The paper is
organized
 as follows. In the next section we list the basic definitions and equations of
 the FCM, related to flux tubes, and in the section 3 define the fields inside
 flux tubes in terms of FC, and magnetic currents. In section 4 our results are
 shown and compared to existing data for $T=0$. Discussion of results and prospectives
 are given in the concluding section.

\section{Field   correlators in QCD}

The vacuum fields $F_{\mu\nu} (x)$ in QCD without  external currents are
necessarily stochastic and can be characterized by the set of Field Correlators
(FC), which in the gauge invariant form for the lowest one, the Gaussian,
\cite{29a,29b,29c} can be written  as

$$  g^2 D^{(2)}_{i4k4} (x-y) \equiv \frac{g^2}{N_c} \lan tr_f (F_{i4} (x) \Phi(x,y) F_{k4} (y)
\Phi(y,x)\ran = (\delta_{ik} ) D^E(x-y)+$$ \be + \frac12 \left(
\frac{\partial}{\partial x_i} [h_k + {\rm ~perm}]\right) D_1^E (x-y), ~~
h_\lambda = x_\lambda -y_x.\label{1}\ee

The temporal Wilson loop in terms of this basic FC can be written as \cite{29a,29b,29c}
via colorelectric FC, $D^E$ and $D^E_1$

$$ W(C) = \frac{1}{N_c} \lan   tr P\exp (ig \int_C dz_\mu A_\mu(z))\ran =$$
$$ \frac{1}{N_c} \left\lan   tr P\exp (ig \int_{S_{\min}} d\sigma_{\mu\nu}
F_{\mu\nu})\right\ran = \frac{1}{N} tr P \exp [-\frac{g^2}{2} \int  d
\sigma_{\mu\nu} d\sigma_{\lambda\rho} \lan F_{\mu\nu} F_{\lambda\rho}\ran
+...]\cong$$
 \be\cong \exp \left( - S_{\min} \frac12 \int D^E (z) d^2z\right),\label{2}\ee
which implies that the string tension is expressed via $D^E(z)$ as \be \sigma^E
= \frac12 \int d^2 z D^E (z).\label{3}\ee

In (\ref{3}) the integration is over the minimal surface $S_{\min}$ inside the
Wilson loop $C$.

Using $D^E$ and $D^E_1$ one can define the instantaneous interaction between
fundamental or adjoint color changes, as it shown in the Appendix 1.

Note that $D^E ( x)$, which enters in (\ref{2}), generates the scalar potential
$V_D(r)$ \be V_D(r) = 2 c_a \int^r_0 (r-\lambda) d\lambda \int^\infty_0 d\nu
D^E (\lambda, \nu) = V_D^{\rm (lin)}(r) + V_D^{\rm sat} (r).\label{4}\ee The FC
$D_1^E$, which enters in the full derivative in (\ref{1}), creates the
vector-like interaction \be V_1 (r) =c_a \int^r_0 \lambda d\lambda
\int^\infty_0 d\nu D^E_1 (\lambda, \nu), ~~ c_{\rm fund} =1, ~~ c_{\rm adj} =
\frac94.\label{5}\ee

Eqs. (\ref{4}) and (\ref{5}) yield the information on correlators $D^E,D^E_1$,
which can be obtained from the study of $Q\bar Q$ potentials $V_D(r),\,V_1(r)$.
In what follows we shall exploit another way: on one side we shall  define
$D^E$ and $D^E_1$ via gluelump Green's function, on  another side we find the
structure of flux tubes with the help of $D^E, D_1^E$. In this way the data on
flux tubes can be  predicted and compared with lattice or experimental sources.

Till now the properties of $D^E$ and $D^E_1$ were not defined and to get
information on that, one should exploit their connection to the gluelump
Green's  function, as it was done in \cite{36a,36b,36c}. Namely, $D^E(x)$ is expressed
via the two-gluon-gluelump Green's function $G^{(2g)} (x,y)$ \be D^E(x-y) =
\frac{g^4 (N_c^2-1)}{2} G^{(2g)} (x,y).\label{6}\ee The lowest eigenvalues and
the asymptotics of $G^{(2g)}$ were found in \cite{34}, namely $M_0^{(2gl)}
\approx 2.5$ GeV,  and from \cite{36a,36b,36c}

\be G^{(2g)} (x) (x\gg 1/M_0^{ (2gl)} ) \approx 0.108 ~\sigma^2_f
e^{-M_0^{(2gl)}|x|}.\label{7}\ee

The insertion of (\ref{7}), (\ref{6}) into (\ref{4}) immediately yields the
potential $V_D^{(\rm lin)}(r)$ which is linear in whole region $r>1/M_0^{(2g)}
\approx 0.1$~fm. This fact agrees well  with all experimental and numerical
data.

It is interesting, that the same approach of two-gluon gluelump asymptotics for
the colormagnetic function $D^H(x-y)$ yields \cite{43} the well-known relation,
found also  on the lattice \cite{44} \be \sigma^H= g^4 (N^2_c-1) T^2 c_\sigma,
~~ c_\sigma = {\rm const}.\label{8}\ee In contrast to that, the FC $D^E_1$ is
expressed via the one-gluon gluelump Green's function with the nonperturbative
part, behaving asymptotically as \cite{36a,36b,36c}

\be D_1^{E({\rm nonp})} (x) = \frac{2N_c \alpha_s}{x} M_0^{(1 gl)}\sigma_f
e^{-M_0^{(1gl)} |x|}\label{9}\ee where $M_0^{(1gl)} \cong  1.5$ GeV
\cite{34,35a,35b}, while the total form, containing the perturbative part as shown
in Appendix 2, is  \be D_1^{E(\rm pert)} (x) =
\frac{2(N_c^2-1)\alpha_sK_2(mx)}{N_c\pi x^2} + O(\alpha^2_s).\label{10}\ee

As a result, the FC $D^E_1$ produces the  interaction $V_1^E(r)$ \be   V_1^{E}
(r) = - \frac{(N^2_c -1)\alpha_se^{-mr}}{2N_c r}+ O(\alpha^2_s),\,\, \text{ where } m=O(1\text{ GeV}).\label{11}\ee

One should stress at this  point, that one-gluon and two-gluon gluelumps enter
separately in their mass measurements both on the lattice  \cite{35a,35b} and
analytically \cite{34}, yielding different values ($2.5$ GeV in (\ref{7}) and $1.5$
GeV in (\ref{9}) in both approaches. On the other hand,  when computing the FC
via gluelumps, as in \cite{36a,36b,36c}, the mixing term appears $(D^{(1)}_{\mu\nu,
\lambda\sigma} $ in \cite{36a,36b,36c}) which mixes up two  contributions and brings in
an averaged value $\bar M$ of the order of $1.0$ GeV. The resulting mass $m$ of the order of $\bar{M}$ in Eq.(\ref{11}) is the gluon screening mass.

At this point we turn to the measurements the flux tube fields, as it is done on the lattice, where one computes the average value of the contour, shown in Fig.1, consisting of a small plaquette at the point $x$, connected by two fundamental lines to the Wilson loop of heavy quarks $Q$, $\bar{Q}$.

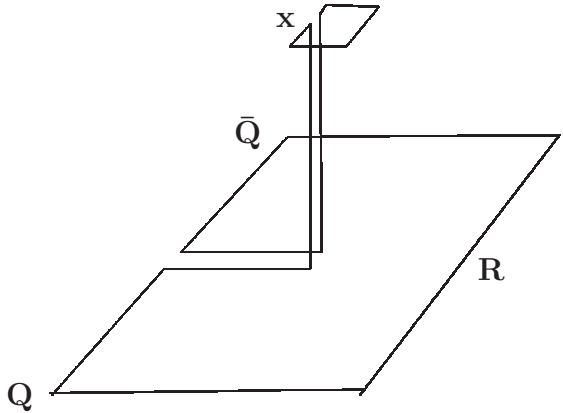
\begin{figure}[h] 
\setlength{\unitlength}{1.0cm}
\centering
\unitlength 1mm 
\linethickness{0.5pt}
\ifx\plotpoint\undefined\newsavebox{\plotpoint}\fi
\begin{picture}(105,90)(0,0)
\multiput(20.25,38.25)(.033675799,.038242009){438}{\line(0,1){.038242009}}
\put(35,55){\line(1,0){19.25}} \put(54.25,55){\line(0,1){32.5}}
\multiput(54.25,87.5)(-.0335366,-.0365854){82}{\line(0,-1){.0365854}}
\put(51.5,84.5){\line(1,0){7.5}}
\multiput(59,84.5)(.03373016,.04166667){126}{\line(0,1){.04166667}}
\multiput(63.25,89.75)(-.875,.03125){8}{\line(-1,0){.875}}
\multiput(56.25,90)(-.032609,-.054348){23}{\line(0,-1){.054348}}
\multiput(55.5,88.75)(.03125,-3.9375){8}{\line(0,-1){3.9375}}
\put(55.75,57.25){\line(-1,0){18.5}}
\multiput(37.25,57.25)(.03373494,.036746988){415}{\line(0,1){.036746988}}
\multiput(51.25,72.5)(4.46875,.03125){8}{\line(1,0){4.46875}}
\multiput(87,72.75)(-.0336970475,-.0442875481){779}{\line(0,-1){.0442875481}}
\multiput(61.25,39)(-2.75,-.033333){15}{\line(-1,0){2.75}}
\put(51,88.0){\makebox(0,0)[cc]{$\mathbf{x}$}} \put(78,55){\makebox(0,0)[cc]{$\mathbf{R}$}}
\put(16,38){\makebox(0,0)[cc]{$\mathbf{Q}$}}
\put(46,73){\makebox(0,0)[cc]{$\mathbf{\bar Q}$}}
\end{picture}
\vspace{-3.5cm}
\caption{The connected probe plaquette at the point $x$ above the $Q\bar{Q}$ Wilson loop.}
\vspace{-0.1cm}
\label{fig:fig01}
\end{figure}
\vspace{0.5cm}

Here appears a new phenomenon, which might be called ``the quenching
of the screening gluon mass'', namely, as shown in Fig.2, the value of
the one-gluon screening mass $M\cong 1.5$ GeV is obtained, when the parallel transporter in the transverse position is fixed, as shown in Fig. 1, where the double fundamental line in the
transverse direction defines the form of the confining film in the gluelump
Green's function.

\begin{figure}[h] 
\setlength{\unitlength}{1.0cm}
\centering
\unitlength 1cm 
\linethickness{0.4pt}
\begin{picture}(5,5)(0,0)
\put(0.5,0.5){\includegraphics[height=4.0cm]{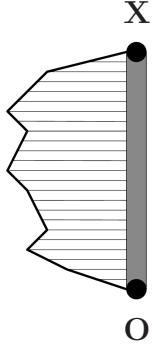}}
\put(2.62,4.5){\makebox(0,0)[cc]{$\mathbf{X}$}} \put(2.62,0.3){\makebox(0,0)[cc]{$\mathbf{O}$}}
\end{picture}
\vspace{-0.1cm}
\caption{The 1g gluelump configuration for the transverse probe.}
\vspace{-0.1cm}
\label{fig:fig02}
\end{figure}
\vspace{1.0cm}

On the other hand, measuring gluon exchange in the  horizontal plane without any transporters, which limit the size of the confining film, one obtains, that the resulting minimal surface is a slightly
deformed plane inside the big contour, as shown in Fig. 3. The energy of
deformation is equal to $\varepsilon_{\rm plane} =\sigma \Delta  S_{\min} =
\sigma \frac{h^2}{L}$, where $h$ is  the average  deflection of gluon path from
the plane and $L$ is plane length. This should be compared with the energy in
the gluelump case, $\varepsilon_{glp} \cong \sigma h$, with the result $
\varepsilon_{\rm plane} \ll \varepsilon_{glp},$ meaning a strong damping of the Coulomb screening $M_{scr}^{Coul}\ll M_{glp}$. 

\begin{figure}[h] 
\setlength{\unitlength}{1.0cm}
\centering
\unitlength 1cm 
\linethickness{0.4pt}
\begin{picture}(10.0,6.0)(0,0)
\put(0.1,0.1){\includegraphics[height=6.0cm]
{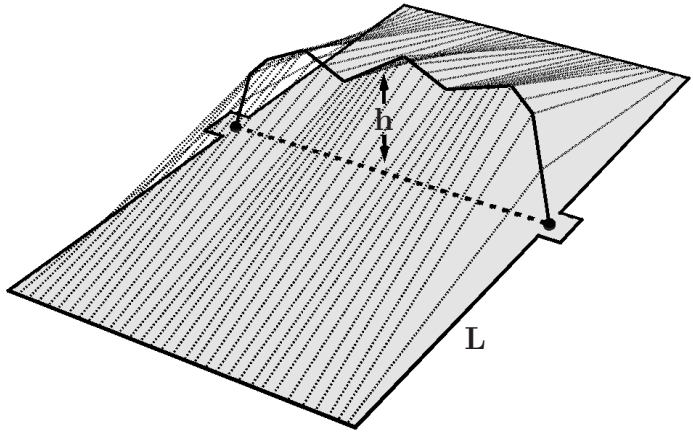}}
\put(5.3,4.4){\makebox(0,0)[cc]{$\mathbf{h}$}} \put(6.5,1.5){\makebox(0,0)[cc]{$\mathbf{L}$}}
\end{picture}
\vspace{-0.1cm}
\caption{The minimal area surface for the gluon exchange interaction.}
\vspace{-0.1cm}
\label{fig:fig03}
\end{figure}\vspace{1.0cm}

Moreover, the length $L$ between consequitive gluon exchanges can be estimated from the action exponent $\operatorname{exp}(-V_{Coul}(R)L)$, as $L^{-1} \sim V_{Coul}(R) \sim \frac{4}{3}\frac{\alpha_s}{R}$, leading to the result $\varepsilon_{plane}=\sigma\frac{h^2}{L}\sim \frac{4}{3}\sigma\frac{h^2}{R}\alpha_s$. Thus the screening is additionally damped at large $R$, for the analytic treatment of this type of interaction (but without $\alpha_s/R$) see \cite{44*}.

It is clear from the analysis of \cite{44*}, that the screening mass, corresponding to the Fig. 3, is the mass excitation of the static hybrid of the length $L$ with the transverse excitation of the order of $\sqrt{12}/L$ and longitudinal $(\sigma/L)^{1/3}$. Defining $L$ as before from the condition $\Delta t\,V_{Coul}=O(1)$, one obtains for $R=1$ fm for both excitations $\Delta m \approx 0.4$ GeV instead of $1.4$ GeV for the one-gluon
gluelump mass. Note also, that for the single gluon exchange, when $\Delta t$ tends to infifnity, the screening mass $\Delta m$ tends to zero.

Numerically on the lattice the static $Q\bar Q$ potential demonstrates
\cite{45a,45b} the linear plus pure Coulomb form. Our discussion above might give an answer to the similar point, raised recently in \cite{46}.

In our case, where a part of parallel transporters is in the transverse
direction, as in Fig. 1 , one expects, that our mass $m$ satisfies
$M^{Coul}_{scr} < m < M_{glp}$, and we choose explicitly in what follows for
transverse $r_\bot$ distributions, $m=1$ GeV,  which we shall use to define the
Vacuum Correlation Length $\lambda$, \be \lambda = \frac{1}{{m}} =
 0.2~{\rm fm}.\label{12}\ee

 To complete the picture of potentials $V_D  $ and $V_1$ one should mention  another important feature of resulting
potentials in (\ref{4}), (\ref{11}): as it is shown in the appendix of
\cite{39}, the terms $V_D^{\rm sat} (r)$ in (\ref{4}) and  $V_1^{(np)} (r)$
have different signs and almost fully compensate each other for low
temperatures. As a result the interaction between two static charges acquires
the well established form, confirmed on the lattice and in experiment: \be V_{Q\bar Q} (r) = V_1^{\rm (pert)}(r) +
V_D^{(\rm lin)}(r).\label{13} \ee

As we shall see below, this cancellation holds only for potentials, which are
in-plane integrals of the FC, as in (\ref{4}), (\ref{5}). However, for  the
flux tube probes, which are mostly the out-of-plane integrals of FC, this full
cancellation does not take place, and one has a possibility of defining the FC
through the measurements of flux tube probes, which is especially interesting
for nonzero $T$, and around $T=T_c$ \cite{47a,47b,48}.

In the next section we define the  connected flux tube probes via field
correlators, following \cite{37,38} and adding a new  contribution from the
correlation $D_1^{(np)}$.

\section{Flux tube fields via field correlators}

 To measure field distributions around the static color charges $Q, \bar Q$,
 one can use the connected probe, defined by the contour $C$, shown in Fig.1,
 as it is done on the lattice \cite{1,2,3,4,5,6,7,8,9,10,11,12,13,14,15,16,17,18,19,20,21,22,23,24}, and calculated in the FCM, see  e.g.
 \cite{37,38}.

 As shown in \cite{38}, Eq.
 ({30}), the resulting effective field $F_{\mu\nu} (x)$ is expressed via the FC, Eq.
 (\ref{1}),
 \be F_{\mu\nu} (x) =\int_S d \sigma_{\alpha\beta} (y) g^2 D^{(2)}_{\alpha\beta
 \mu\nu} (x-y).\label{14}\ee

 Inserting (\ref{1}) in (\ref{14}) one obtains as in \cite{38} the expression
 for the colorelectric probe $\veE_i (\ver, \veR)$

 \be E_i (\ver, \veR) = n_k \int^R_0 dl \int^\infty_{-\infty} dt \left(
 \delta_{ik} D^E (z) + \frac12 \frac{\partial (z_iD_1^E(z))}{\partial z_k}
 \right), \label{15}\ee
where $z=(\ver - \ven l,t),$ and $ \ven = \frac{\veR}{R}$ is along the axis
$x_3$, where the charges $Q, \bar Q$ are placed at the distance $R$, see Fig.
1.

Inserting the perturbative part $D^{E(\rm pert)}_1$ from (\ref{10}), one
obtains the screened  color Coulomb field   \be \veE^{(1)} (\ver) =
\frac{(N^2_c-1)}{2N_c} \left( \frac{ \alpha_s \ver \chi (mr)}{r^3} - \frac{
\alpha_s (\ver- \veR)}{(\ver-\veR)^3}\chi (m|\ver-\veR|)\right),\label{16}\ee
where the screening factor is $\chi (z) = (1+z) e^{-z}$,  and at the midpoint
between the charges one has \be \veE^{(1)} \left( \frac{\veR}{2}\right) =
\frac{4(N^2_c-1)}{N_c} \frac{\alpha_s \veR}{R^3}\chi \left(
\frac{mR}{2}\right).\label{17}\ee

Eq.(\ref{17}) contains both the standard perturbative part $E^{(1)}\sim
\frac{\alpha_s}{R^2},$ at $ R\ll \frac{1}{m}$ and the nonperturbative screening
$E^{(1)} \sim \frac{\alpha_s m}{R} e^{-mR}$ at $R\gg \frac{1}{m}$. For the
field correlators, as in (\ref{14}), the mass, $m\approx 1$ GeV, while the
screening in the OGE potential $V_1 (r)$ Eq. (\ref{11}), is much softer, as
discussed in \cite{49}, see also appendix 2.

 From (\ref{17}) one can estimate
$E_3^{(1)} $ at $R=0.2$ fm, $m=1$ GeV and $r_{\bot} =0$. 
$E_3^{(1)}(0.5\text{ fm},\,0)=0.122\alpha_s$ GeV$^2\approx 0.05$ GeV$^2$ for $\alpha_s \cong 0.4$.
 In a similar way, using the asymptotics (\ref{7}), and the relation (\ref{3}),
 one has
 \be D(z) = \frac{\sigma}{\pi \lambda^2}\exp \left( - \frac{
 |z|}{\lambda }\right), ~~ \lambda^{-1}  \cong 1 ~{\rm GeV} ,\label{21}\ee
 which yields for the colorelectric probe, following (\ref{15})
\be \veE^D = \ven \frac{2\sigma}{\pi} \int^{R/\lambda}_0 dl \left| l\ven -
\frac{\ver}{\lambda}\right| K_1 \left( \left| l\ven-
\frac{\ver}{\lambda}\right|\right).\label{22}\ee

For $R\to \infty$ one obtains from (\ref{22}) the saturated  colorelectric
field at the distance $r_\bot$ from the axis \be E_3^D (\ver_\bot) = 2 \sigma
\left( 1+ \frac{r_\bot}{\lambda} \right) \exp \left( -
\frac{r_\bot}{\lambda}\right)\label{23}\ee and the saturated on-axis value
$E^{\rm sat}_3 $(on axis) = 2$\sigma$.

Summing up the contributions of (\ref{17}), (\ref{22})  for the field $E_3$ at
the midpoint on the axis $(r_\bot =0)$, one has $(N_c=3)$ \be E^{\rm tot}_3
\left( \frac{R}{2}, r_\bot=0 \right) = \frac{32\alpha_s}{3R^2}\,\,\chi\left(
\frac{mR}{2}\right)   + \frac{2\sigma}{\pi} \int^{R/\lambda}_0 dx\cdot x K_1
(x).\label{24}\ee

Another interesting characteristics of flux tabes is the $E_3$ dependence on
the  distance to the $Q\bar Q$ axis, i.e. on $r_\bot$. Using (\ref{16}),
(\ref{22})  one can write

\be E_3(r_\bot) \equiv E_3 \left(r_\bot, \frac{R}{2}\right) = E^D_3
\left(r_\bot, \frac{ R}{2}\right) + E^{(1)}_3 \left(r_\bot, \frac{ R}{2}\right)
, \label{25}\ee where \be E_3^D\left(r_\bot, \frac{ R}{2}\right) =
\frac{2\sigma}{\pi} \int^{\frac{R}{2\lambda}}_{-\frac{R}{2\lambda}} dx \sqrt{
x^2 + \frac{r^2_\bot}{\lambda^2}} K_1\left(\sqrt{ x^2 +
\frac{r^2_\bot}{\lambda^2}}\right),\label{26}\ee $E^{(1)}_3$ is given in
(\ref{16}),  \be E^{(1)}_3\left(r_\bot, \frac{R}{2}\right)=\frac{4}{3}
\alpha_s \frac{R\chi(m \sqrt{r^2_\bot+ \frac{R^2}{4}})}{\left(r_\bot^2+
\frac{R^2}{4}\right)^{3/2}}.\label{27}\ee

In the next  section we compare our results for $E_3(r_\bot)$ and $E^{tot}_3
\left(\frac{R}{2}\right)$ with the lattice data \cite{40}.

We now turn to the  effective magnetic monopole picture, which can be derived
from our method, to  compare it with the dual superconductor model.

To this end we as in \cite{38} define first of all the magnetic current $\vek$,
\be  \vek = {\rm rot}\,\,\veE (\ver, \veR) ={\rm rot}\,(\veE^D (\ver)+ \veE^{(1)}
(\ver))\equiv \vek_D +\vek^{(1)},\label{28}\ee

One can deduce from (\ref{16}), that $\veE^{(1)}$ at  $ r_3 = \frac{R}{2}$ (at
the midpoint) does not have component  along axis 1 and 2, so that it can be
written as $\veE^{(1 )} (r_3=\frac{R}{2}, r_\bot)= \ven f^{(1)} (r^2_\bot)$,
and hence $f^{(1)} (r^2_\bot) = E_3^{(1)} (r_\bot, \frac{R}{2}) $ given in
(\ref{25}).

The same is true  for $E^D$,  Eq. (\ref{22}), so that   the total $\vek=
\vek^{(1)} +\vek_D$ is perpendicular to $\ven$, as shown in Fig. 4.

\begin{figure}[h] 
\setlength{\unitlength}{1.0cm}
\centering
\unitlength 1cm 
\linethickness{0.4pt}
\vspace{0.5cm}
\begin{picture}(7.0,5.0)(0,0)
\put(0.5,0.5){\includegraphics[height=5.0cm]
{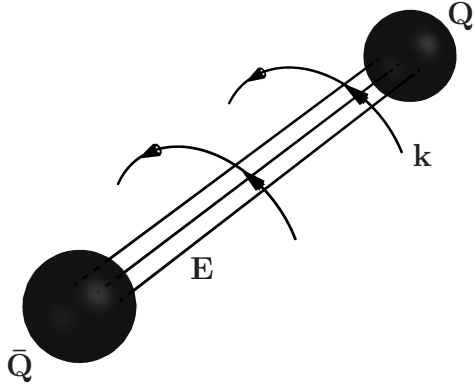}}
\put(6.4,5.25){\makebox(0,0)[cc]{$\mathbf{Q}$}}
\put(0.6,0.6){\makebox(0,0)[cc]{$\mathbf{\bar Q}$}}
\put(5.9,3.4){\makebox(0,0)[cc]{$\mathbf{k}$}}
\put(3.0,1.9){\makebox(0,0)[cc]{$\mathbf{E}$}}
\end{picture}
\vspace{-0.1cm}
\caption{Colormagnetic current $k$ around the flux tube.}
\vspace{-0.1cm}
\label{fig:fig04}
\end{figure}\vspace{1.0cm}

Therefore $k_z =0$, and $k_x =2y f', k_y=-2xf'$ so that $\vek\ver =0$, which
means, that vectors $\vek$  form  circular loops around the $(Q,\bar Q)$ axis.
The  function $f^{(1)} (r_\bot^2)=E_3^{(1)}$ is given in (\ref{27}), and one
can calculate the  $r_\bot$ dependence  $|\vek_\bot^{(1)} (r_\bot)|^2 = (
k_x^{(1)} (r_\bot))^2+( k_y^{(2)} (r_\bot))^2, $

\be (\vek^{(1)} (r_\bot))^2  =4r^2_\bot \left( \frac{\partial
E^{(1)}_3}{\partial r^2_\bot}\right)^2
 .\label{29}\ee 

The function $\vek_D (r_\bot)$  is obtained from (\ref{26}), where one can use
the relation $\frac{d}{dz} (zK_1(z)) = - z K_0 (z)$, with the result

\be \vek^2_D (r_\bot) = \frac{4\sigma^2 r^2}{\pi^2 \lambda^4}  \left(
\int^{\frac{R}{2\lambda}}_{-\frac{R}{2\lambda}} dx K_0\left(\sqrt{ x^2 +
\frac{r^2_\bot}{\lambda^2}}\right)\right)^2\label{30}\ee

In the case,  when $\vek_D$ and $\vek^{(1)}$ can be both nonzero the resulting
$\vek(r_\bot)$ is

\be\vek  (r_\bot) = \vek^{(1)}+\vek_D=- (\ven)_\varphi r_\bot   \left(
\frac{2\sigma}{\pi\lambda^2} \int^{\frac{R}{2\lambda}}_{-\frac{R}{2\lambda}} dx
K_0\left(\sqrt{ x^2 + \frac{r^2_\bot}{\lambda^2}}\right)+ \frac{\partial\,f^{(1)}}{\partial\,r^2_\bot}\right)\label{31}\ee and $\ven_\varphi \ver_\bot =0,\,\,\ven^2_\varphi =1$.

The most important point for the connection to the superconducting  model is
the dual London equation ${\rm rot}\,\vek = \lambda^{-2} \veE$, which,    as
shown in \cite{38}, is supported asymptotically $(r_\bot \to \infty)$ by the
relation for the saturated string $( R\to \infty)$ \be {\rm rot}~ \vek_D =
\gamma_D (r_\bot) \lambda^{-2} \veE^{(D)} (r_\bot), ~~ \gamma_D(r_\bot) =
\frac{r_\bot/\lambda-2}{r_\bot/\lambda+1}, \gamma(\infty) =1.\label{32}\ee 

\section{Results and discussion}

To compare with recent accurate lattice data \cite{40}, we are using the data, shown in Fig. 4 of this paper for two types of behavior: first, we are using the data \cite{40} for $E_3^{tot}(R)$ here $R=0.76$ fm, $0.95$ fm, $1.14$ fm, $1.33$ fm and calculate our $E_3$ from Eq.(\ref{21}) for these values of $R$. The results are shown in Fig. 5 with $\alpha_s=0.4$, $m=1$ GeV, $\sigma=0.18$ GeV$^2$. One can see a reasonable agreement of our theory with the data,where a slow decrease of $E_3^{tot}(R)$ is due to $E_3^{(1)}$, while the saturation at $E_3=2\sigma$ is due to $E_3^D$.

\begin{figure}[h]
\setlength{\unitlength}{1.0cm}
\vspace{0.5cm}
\begin{center}
\begin{picture}(10.0,8.0)(0,0)
\put(0.5,0.5){\includegraphics[height=8.0cm]{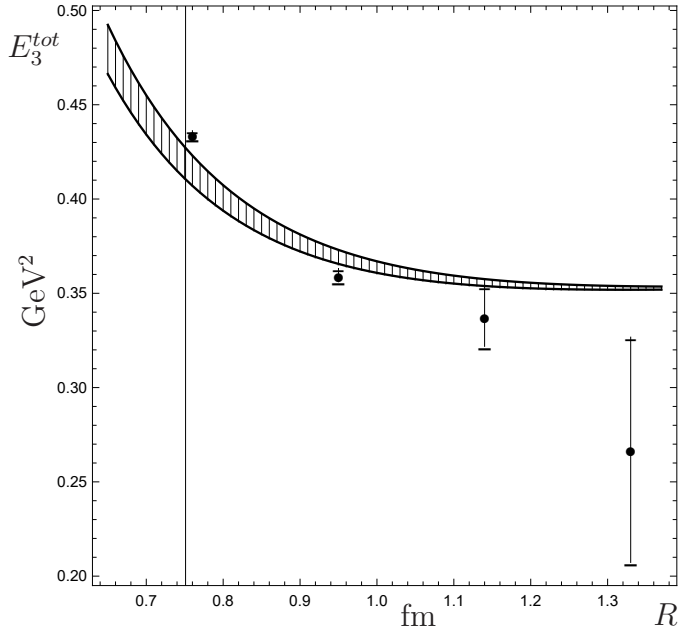}}
\put(0.2,7.9){\makebox(0,0)[cc]{{$E_3^{tot}$}}}
\put(8.5,0.4){\makebox(0,0)[cc]{$R$}}
\put(0.0,4.2){\rotatebox{90}{GeV$^2$}}
\put(5.25,0.4){\makebox(0,0)[cc]{fm}}
\end{picture}
\caption{$E_3^{tot}=E_3^{tot}\left( \frac{R}{2}, r_{\perp}=0 \right)$. The tube length dependence of the CE field strength at the midpoint. The shaded region corresponds to $\alpha_s=0.4$ (lower curve) and $\alpha_s=0.45$ (upper curve). The points with error bars are from the lattice measurements in \cite{40}.}
\label{fig:fig05}
\end{center}
\end{figure}

To check the $r_\perp$ dependence, we again are using data of \cite{40} and present our results for the quoted values of $R$ in Figs. 6-9. One can see again a reasonable agreement at the level of $O(5\%)$ for $r_{\perp}<0.5$ fm. Note, that our  parameters $\alpha_s$, $m$, $\sigma$ are fixed at the physically relevant values, $\alpha_s$($Q\sim$~$1$~GeV)$\,=0.4$, $\sigma=0.18$ GeV$^2$.

\begin{figure}[h]
\setlength{\unitlength}{1.0cm}
\begin{center}
\begin{minipage}[h]{0.48\linewidth}
\begin{picture}(7.25,7.0)
\put(0.3,0.3){\includegraphics[height=6.85cm]{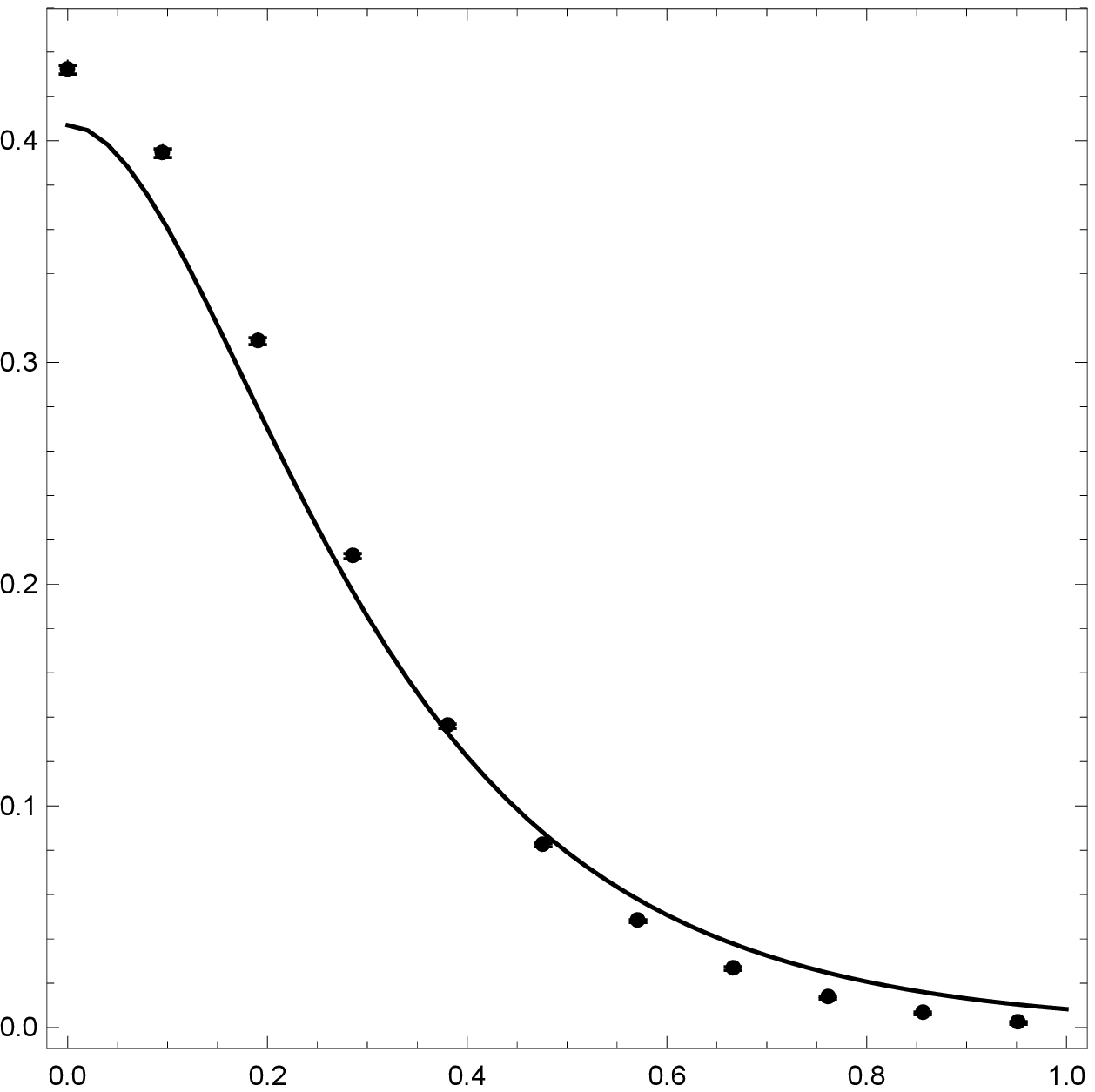}}
\put(0.25,7.0){\makebox(0,0)[cc]{{$E_3$}}}
\put(7.2,0.0){\makebox(0,0)[cc]{$r_{\perp}$}}

\put(-0.1,3.3){\rotatebox{90}{GeV$^2$}}
\put(3.95,0.15){\makebox(0,0)[cc]{fm}}
\end{picture}
\caption{$E_{3}=E_{3}(r_{\perp},\,R=0.76\text{ fm})$. The transverse radius dependence of the CE field strength for the fixed flux tube length $R=0.76\text{ fm}$. The dots with error bars are from the lattice measurements in \cite{40}.}
\label{fig:fig06}
\end{minipage}
\hfill
\begin{minipage}[h]{0.48\linewidth}
\begin{picture}(7.25,7.0)
\put(0.3,0.3){\includegraphics[height=6.85cm]{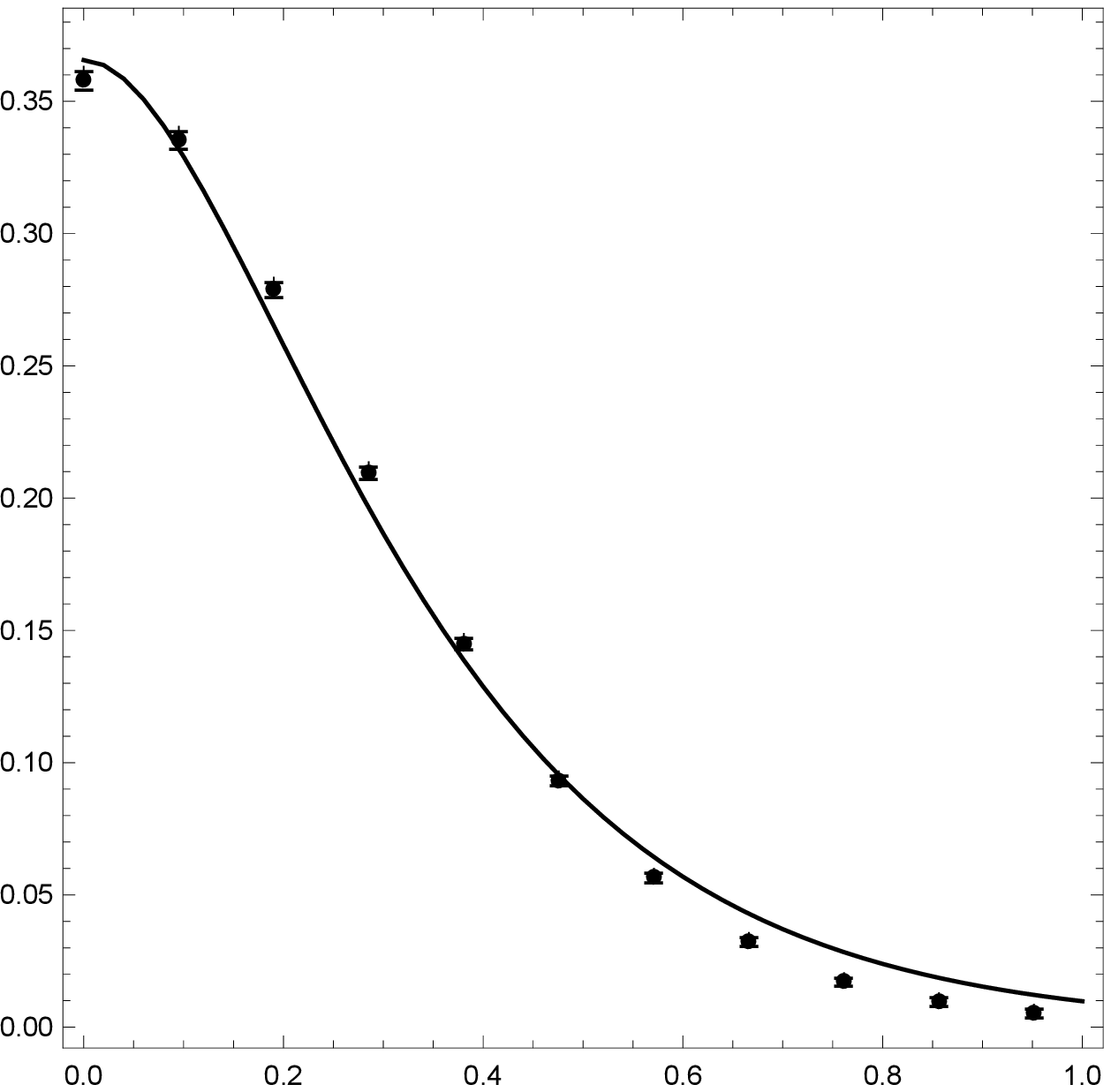}}
\put(0.25,7.0){\makebox(0,0)[cc]{{$E_3$}}}
\put(7.2,0.0){\makebox(0,0)[cc]{$r_{\perp}$}}

\put(-0.1,3.3){\rotatebox{90}{GeV$^2$}}
\put(3.95,0.15){\makebox(0,0)[cc]{fm}}
\end{picture}
\caption{$E_{3}=E_{3}(r_{\perp},\,R=0.95\text{ fm})$. The transverse radius dependence of the CE field strength for the fixed flux tube length $R=0.95\text{ fm}$. The dots with error bars are from the lattice measurements in \cite{40}.}
\label{fig:fig07}
\end{minipage}
\end{center}
\end{figure}

\begin{figure}[h]
\setlength{\unitlength}{1.0cm}
\begin{center}
\begin{minipage}[h]{0.485\linewidth}
\begin{picture}(7.25,7.0)
\put(0.3,0.3){\includegraphics[height=6.85cm]{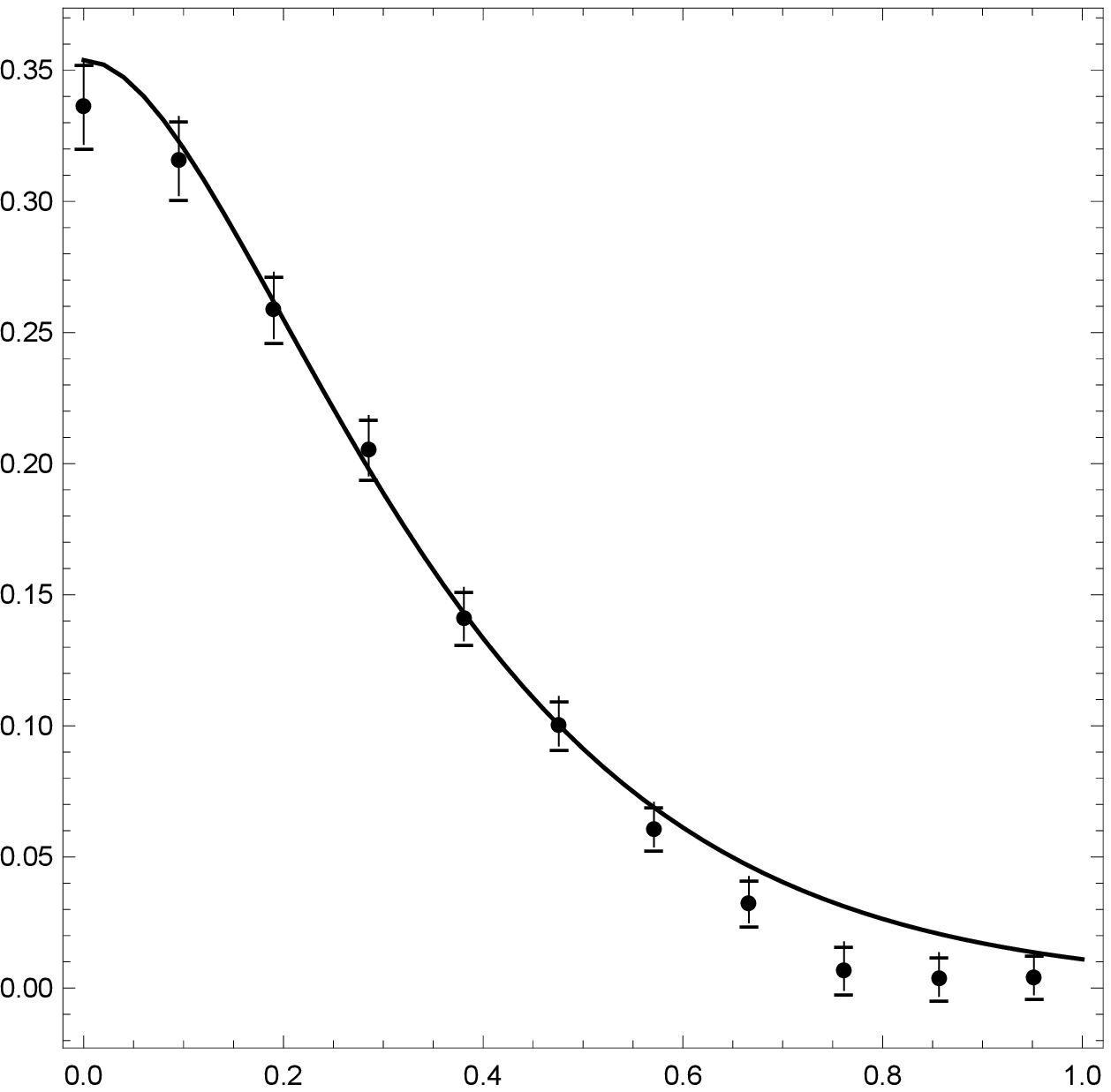}}
\put(0.3,7.2){\makebox(0,0)[cc]{{$E_3$}}}
\put(7.2,0.0){\makebox(0,0)[cc]{$r_{\perp}$}}

\put(-0.1,3.3){\rotatebox{90}{GeV$^2$}}
\put(3.95,0.15){\makebox(0,0)[cc]{fm}}
\end{picture}
\caption{$E_{3}=E_{3}(r_{\perp},\,R=1.14\text{ fm})$. The transverse radius dependence of the CE field strength for the fixed flux tube length $R=1.14\text{ fm}$. The dots with error bars are from the lattice measurements in \cite{40}.}
\label{fig:fig08}
\end{minipage}
\hfill
\begin{minipage}[h]{0.485\linewidth}
\begin{picture}(7.25,7.0)
\put(0.3,0.3){\includegraphics[height=6.85cm]{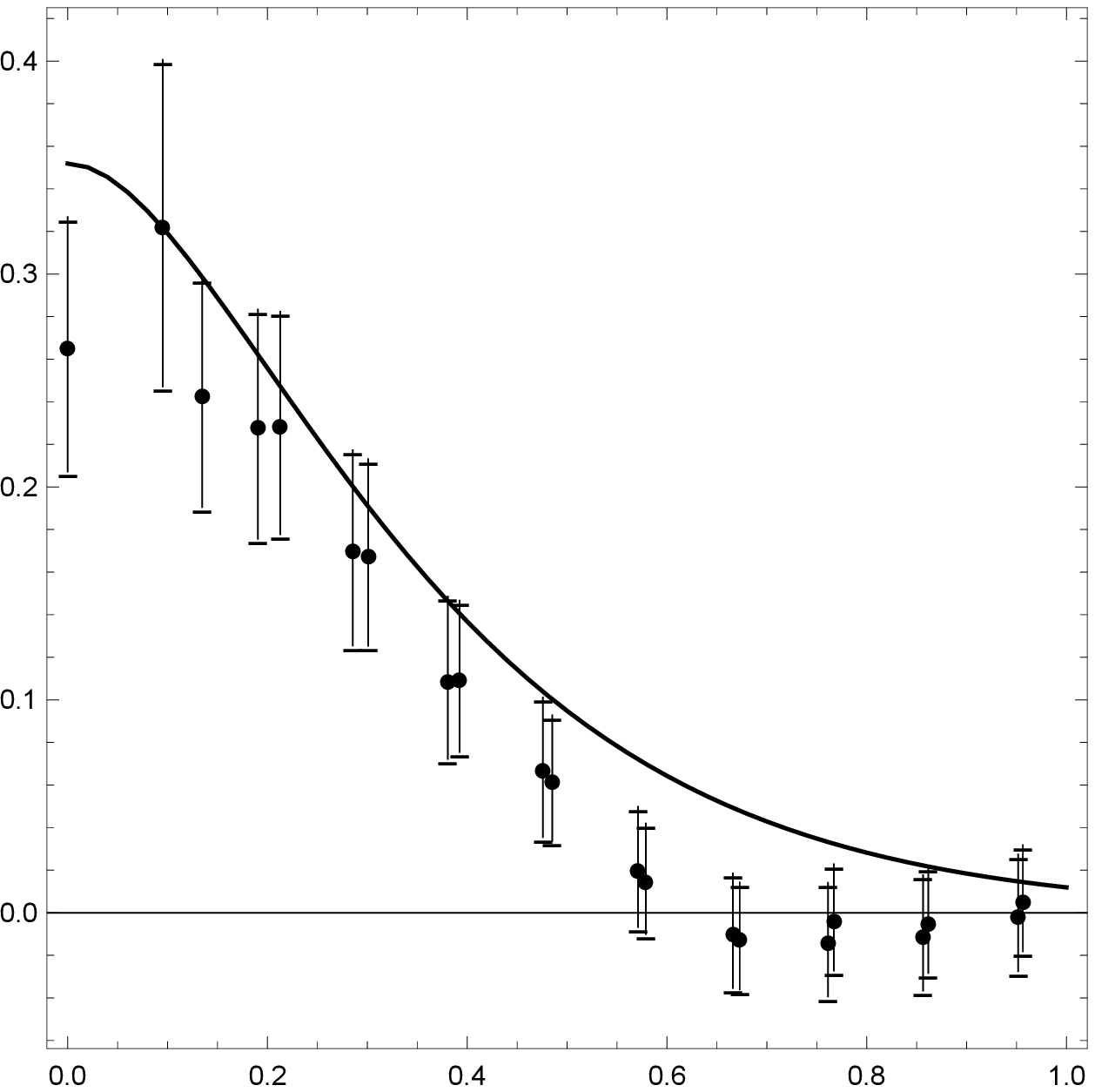}}
\put(0.3,7.2){\makebox(0,0)[cc]{{$E_3$}}}
\put(7.2,0.0){\makebox(0,0)[cc]{$r_{\perp}$}}

\put(-0.1,3.3){\rotatebox{90}{GeV$^2$}}
\put(3.95,0.15){\makebox(0,0)[cc]{fm}}
\end{picture}
\caption{$E_{3}=E_{3}(r_{\perp},\,R=1.33\text{ fm})$. The transverse raius dependence of the CE field strength for the fixed flux tube length $R=1.33\text{ fm}$. The dots with error bars are from the lattice measurements in \cite{40}.}
\label{fig:fig09}
\end{minipage}
\end{center}
\end{figure}

Finally,in Fig. 10 we demonstrate the $r_\perp$ dependence of the modulus of $|\mathbf{k}(r_{\perp})|$, given by Eqs. (\ref{29}),(\ref{31}). One can see the exponential decay at large $r_{\perp}$ ,typical for the color screening of massive gluon fields.

\begin{figure}[h]
\setlength{\unitlength}{1.0cm}
\begin{center}
\begin{picture}(9.2,8.5)
\put(0.3,0.3){\includegraphics[height=7.85cm]{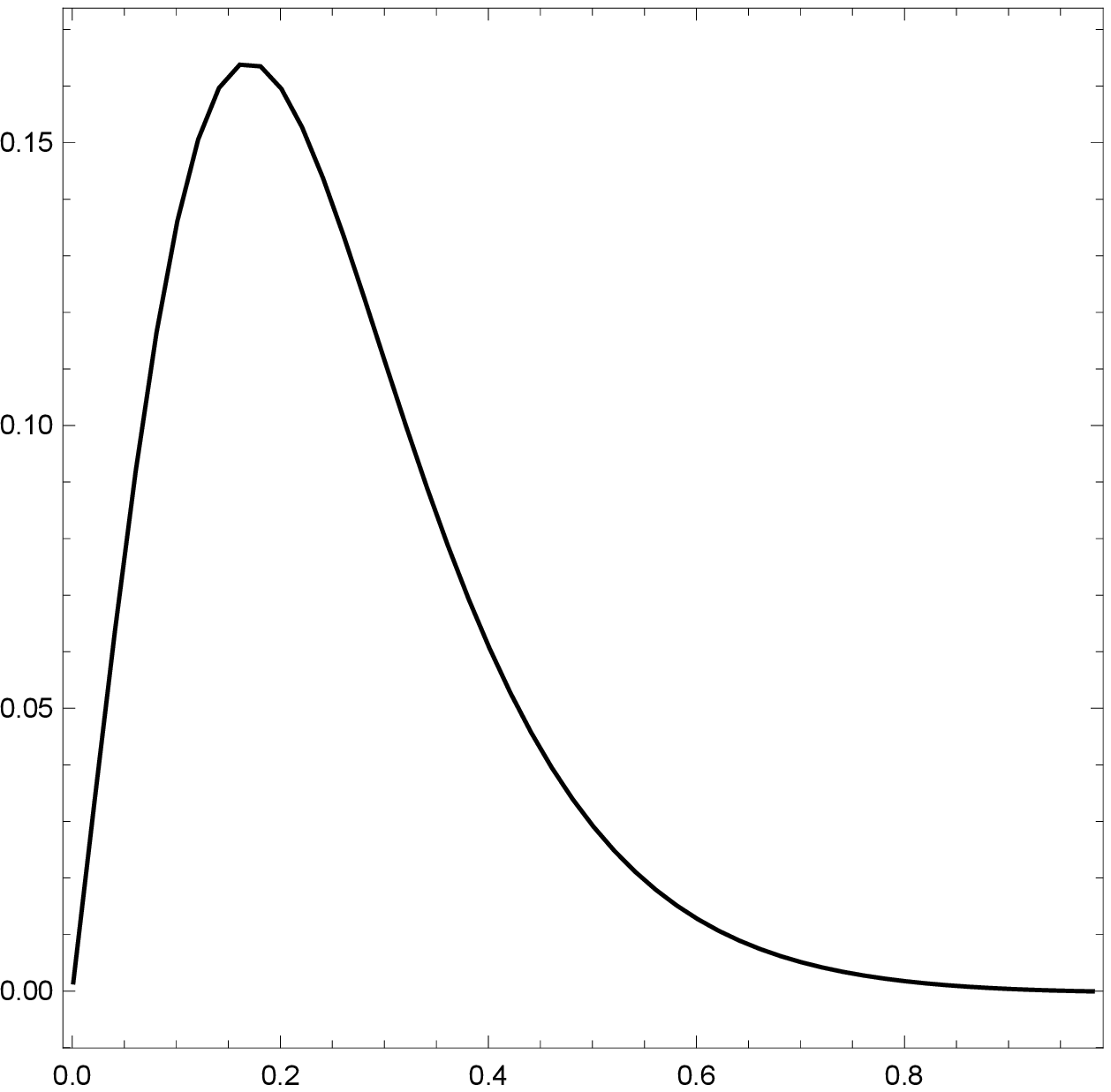}}
\put(0.5,7.95){\makebox(0,0)[cc]{{$\mathbf{k}$}}}
\put(8.2,0.2){\makebox(0,0)[cc]{$r_{\perp}$}}

\put(0.0,4.0){\rotatebox{90}{GeV$^3$}}
\put(4.6,0.2){\makebox(0,0)[cc]{fm}}
\end{picture}
\caption{The transverse radius dependence of the CM current $|{\bf k}(r_{\perp})|$ at $R=0.76\text{ fm}$.}
\label{fig:fig10}
\end{center}
\end{figure}

At this point one can compare our results with the MDS picture. In general, one can treat the dual abelian Higgs picture and different versions of MDS in the same language as in FC, calculating $D^E$, $D^E_1$ via solutions of Ginzburg-Landau equations etc.

This type of analysis was done in \cite{49}, where $D^E$, $D^E_1$ have been related with the dual filed propagator in the abelian Higgs model. However, parameters of the model and the form of $D$, $D_1$ are not fixed, e.g the connection of sigma and the mass $m=1/\lambda$, whereas in the FC approach in QCD the product $\sigma\,\lambda^2$ is fixed by the gluelump mechanism and ensures Casimi scaling, observed on the lattice. Summarising, in the FC approach all observables are defined by the only nonperurbative scale (in addition to quark masses), which can be chosen as $\sigma$.

In \cite{40} the theoretical form of $E_3(r_\perp)$ was chosen, according to the solutions of the Ginzburg-Landau equations, suggested in \cite{50}, with parameters, ensuring a good agreement with the lattice data. These parameters correspond to the superconductor of the first order, where the coherence length $\xi$ is larger, than the penetration length $\lambda$. However, the three flux tube parameters depend (moderately) on the length of the flux tube $R$.

Summarizing,we have derived two components of the CE fields in the flux tube and have shown the strong trasverse screening of CE fields on the length $\lambda=0.2$ fm. We also found the slight decrease and saturation of the on-axis field $E_3(R)$ at large $R$. We have found a reasonably good agreement of our results with the latest lattice data of \cite{40}, confirming the applicablity of our theory using standard parameters, independent of $R$.

Finally, we have presented arguments, why the in-plane  screening of the gluon exchange (color Coulomb interaction) is strongly damped, as compared with the
transverse screening of the same interaction.

The authors are grateful to P.~Cea, L.~Cosmai, F.~Cuteri and A.~Papa for providing the numerical data.

This work was done in the framework of the scientific project, supported by the Russian Science Foundation grant \#16-12-10414.

\vspace{0.4cm}
\setcounter{equation}{0}
\renewcommand{\theequation}{A.\arabic{equation}}

\appendix
\section{Appendix: Calculation of the correlator $D_1$ via the gluelump Green's function} 

\vspace{0.1cm}
\setcounter{equation}{0} 
\def\theequation{A.\arabic{equation}}

Consider the field correlator  Eq.(\ref{1}), and take into account, that
$F_{\mu\nu} = \partial_\mu A_\nu-\partial_\nu A_\mu - ig [A_\mu A_\nu].$ The
contribution of the first terms with derivatives immediately yields the lowest
contribution in the form \be D_1^E (x) = - \frac{2g^2}{N^2_c}
\frac{dG(x)}{dx^2}, \label{A2.1}\ee where $G(x)$ is the one-gluon gluelump
Green's function \be G_{\mu\nu}^{(1g)} (x,y) = \lan Tr_a A_\mu (x) \hat \Phi
(x,y) A_\nu (y) \ran = \delta_{\mu\nu} G(x-y)\label{A2.2}\ee and  $\hat
\Phi(x,y)$ is the parallel transporter in the adjoint representation and we
have exploited the Feynman gauge.

To simplify the matter we consider the gluelump Green's function as a
relativistic Green's function of scalar particle with mass $m$ (neglecting
internal degrees of freedom in the first approximation), which yields \be G(x)
= \frac{(N_c^2-1) N_c}{4\pi^2} \frac{ m}{|x|} K_1 (m|x|),\label{A2.3}\ee where
$K_1$ is the modified Bossel function. Taking derivative  in (\ref{A2.1}), one
has \be D^E_1 (x) = \frac{g^2 m^2 }{4\pi^2} \frac{(N^2_c-1)}{N_c} \frac{ K_2
(m|x|)}{x^2}.\label{A2.4}\ee

In the limit $m\to 0$ Eq. (\ref{A2.4}) yields the standard one-gluon form
$D_1^E (x) = \frac{16\alpha_s}{3\pi x^4}$, which generates according to (5) the
color Coulomb interaction $V_1(r) =- \frac{4\alpha_s}{3r}$.

In the paper the form (\ref{A2.4})  is  used to predict the field distribution
in the flux tube.



\end{document}